\documentclass[11pt]{article}

\usepackage{amssymb,amsmath}
\usepackage{graphicx}
\usepackage{epsfig}
\usepackage{tikz}
\usepackage{color}

\usepackage[english]{babel}

\usepackage{algorithm}
\RequirePackage{natbib}

\definecolor{verm}{rgb}{0.6,0.2,0.2}
\definecolor{purp}{rgb}{0.3,0.1,0.6}
\definecolor{purple}{rgb}{0.4,0.0,0.6}
\definecolor{bggreen}{rgb}{0.1,0.3,0.1}
\definecolor{dgreen}{rgb}{0.1,0.6,0.1}
\definecolor{black}{rgb}{0.0,0.0,0.0}
\definecolor{crim}{rgb}{0.3,0.1,0.1}
\definecolor{dred}{rgb}{0.5,0.1,0.1}

\def\eb{{\bf e}}

\def\B{{\cal B}}
\def\C{{\cal C}}

\def\G{{\cal G}}
\def\H{{\cal H}}

\def\M{{\cal M}}
\def\N{{\cal N}}

\def\R{{\mathbb R}}

\def\al{\alpha}
\def\d{\delta}

\def\e{\epsilon}
\def\g{\gamma}

\def\l{\lambda}
\def\L{\Lambda}

\def\r{\rho}
\def\s{\sigma}
\def\SI{\Sigma}

\def\th{\theta}

\def\ap{\rightarrow}

\def\es{\emptyset}
\def\seq{\subseteq}

\def\bi{\{0,1\}}

\def\bp{\{-1,1\}}
\def\bz{{\bf 0}}
\def\imp{\; \Rightarrow \;}

\def\fa{\; \forall}

\def\st{\mbox{ s.t. }}
\def\sg{\mbox{sign}}
\def\nm{\Vert}

\renewcommand{\and}{\mbox{$\wedge$}}

\newcommand{\bc}{\begin{center}}
\newcommand{\ec}{\end{center}}
\newcommand{\be}{\begin{equation}}
\newcommand{\ee}{\end{equation}}
\newcommand{\bd}{\begin{displaymath}}
\newcommand{\ed}{\end{displaymath}}
\newcommand{\ba}{\begin{array}}
\newcommand{\ea}{\end{array}}
\newcommand{\ben}{\begin{enumerate}}
\newcommand{\een}{\end{enumerate}}
\newcommand{\bit}{\begin{itemize}}
\newcommand{\eit}{\end{itemize}}
\newcommand{\beq}{\begin{eqnarray}}
\newcommand{\eeq}{\end{eqnarray}}
\newcommand{\btab}{\begin{tabular}}
\newcommand{\etab}{\end{tabular}}
\newcommand{\bfig}{\begin{figure}}
\newcommand{\efig}{\end{figure}}
\newcommand{\btp}{\begin{tikzpicture}}
\newcommand{\etp}{\end{tikzpicture}}

\def\cons{{\mbox{const}}}

\def\Rcal{{\cal R}}

\newcommand{\supp}{\mbox{supp}}
\newcommand{\argmin}{\operatornamewithlimits{argmin}}

\def\xh{\hat{x}}

\def\xl{x_{\L}}
\def\xlc{x_{\L^c}}

\def\xloc{x_{\L_0^c}}

\newcommand{\nmeu}[1]{ \nm #1 \nm_2 }

\newcommand{\nmP}[1]{ \nm #1 \nm_P }

\def\ru{\underline{\r}}
\def\rb{\bar{\r}}
\def\GkS{{\rm GkS}}

\newtheorem{theorem}{{\bf Theorem}}[section]

\newtheorem{corollary}{{\bf Corollary}}[section]
\newtheorem{definition}{{\bf Definition}}[section]

\begin{document}

\title{
Machine Learning Methods in the \\ Computational Biology of Cancer
}
\author{M.\ Vidyasagar
\thanks{
Erik Jonsson School of Engineering and Computer Sciences
University of Texas at Dallas
800 W.\ Campbell Road, Richardson, TX 75080, U.S.A.
M.Vidyasagar@utdallas.edu.
This research was supported by National Science Foundation
Awards \#ECCS-1001643 and \#ECCS-1306630,
the Cecil \& Ida Green Endowment at the UT Dallas,
and by a Developmental Award from the Harold Simmons
Comprehensive Cancer Center, UT Southwestern Medical Center, Dallas.}
}

\maketitle

\begin{abstract}

The objectives of this ``perspective'' paper are to review some recent
advances in sparse feature selection for regression and classification,
as well as compressed sensing,
and to discuss how these might be used to develop tools to advance
personalized cancer therapy.
As an illustration of the possibilities,
a new algorithm for sparse regression is presented, and is applied
to predict the time to tumor recurrence in ovarian cancer.
A  new algorithm for sparse feature selection in classification problems
is presented, and its validation in endometrial cancer is briefly discussed.
Some open problems are also presented.

\end{abstract}

\section{Introduction}\label{sec:1}

The objectives of this ``perspective'' paper are to review some recent
advances in sparse feature selection for regression and classification,
and to discuss how these might be used in the computational biology of cancer.

Cancer is the second leading cause of death in the United States
(\cite{SEER}).
It is estimated that in the USA in 2013, there will be
1,660,290 new cases of cancer in all sites, and 589,350 deaths
(\cite{Cancer-Stats13}).
In the UK, in 2011 there were 331,487 cases of cancer, and 159,178 deaths.
both are the latest figures available
(\cite{Cancer-UK}).
Worldwide, cancer led to about 7.6 million deaths in 2008
(\cite{WHO12}).
It is interesting to note that, whether in developed countries such as
the USA and the UK, or worldwide, cancer accounts for roughly 13\% of
all deaths (\cite{WHO12}).

One of the major challenges faced by cancer researchers is that
no two manifestations of cancer are alike, even when they occur
in the same site.
One can paraphrase the opening sentence of Leo Tolstoy's {\it Anna 
Karenina\/} and say that ``Normal cells are all alike.
Every malignant cell is malignant in its own way.''
Therefore when it comes to tackling cancer, it is essential to 
group these multiple manifestations into classes that are broadly
similar within each class and substantially different between classes.
Then attempts can be made to develop therapeutic regimens 
that are tailored for each class.
Though this approach is often referred to in the literature as
``personal'' or ``personalized'' medicine, such nomenclature would be 
optimistic.
It would be more accurate to describe this approach as ``patient
stratification.''
We are quite far away from truly personalized therapy at the level of
a single individual.
However, patient stratification is well within reach.

Until recently, grouping of cancers has been attempted first through the
site of the cancer, and then through
histological considerations, that is, the physical appearance of the tumor,
and other parameters that can be measured by physical examination of the tumor.
During the past decade, attempts have been made to collect the
experimental data generated by various research laboratories into central
repositories such as the Gene Expression Omnibus (\cite{GEO})
and the Catalogue of Somatic Mutations in Cancer COSMIC (\cite{COSMIC}).
However, the data in these repositories is often collected under
widely varying experimental conditions.
To mitigate this problem,
there are now some massive public projects under way for
generating vast amounts of data for all the tumors that are
available in various tumor banks, using standardized sets of
experimental protocols.
Amongst the most ambitious are The Cancer Genome Atlas, usually referred
to by the acronym TCGA (\cite{TCGA})
which is undertaken by the National Cancer Institute
(NCI), and the International Cancer Genome Consortium, referred to also as
ICGC (\cite{ICGC}), which is a multi-country effort.
In the TCGA data, molecular measurements are available for almost all tumors,
and clinical annotations are also available for many tumors.
With such a wealth of data becoming freely available, 
researchers in the machine learning community can now aspire to make
useful contributions to cancer biology 
without the need to undertake any experimentation themselves.
Of course, without close interactions with one or more biologists, such work
would be in a vacuum and have little impact.

One of the motivations for writing this paper is to present 
a broad picture of some recent advances in machine learning to
the more mathematically inclined within the cancer biologist community,
and to apply some of these techniques to a couple of problems.
Full expositions of these applications will be presented elsewhere.
In the other direction, it is hoped that the paper will also
facilitate the entry
of interested researchers from the machine learning community 
into cancer biology.
In order to understand the {\it computational\/} aspects of the
problems described here, a basic
grasp of molecular biology is sufficient, as can be obtained from
standard references, for example
\cite{Northrop-Connor09,Tozeren-Byers03}.

Now we briefly state the class of problems under discussion in this paper.
This also serves to define the notation used throughout.
Let $m$ denote the number of tumor samples that are analyzed,
and let $n$ denote the number of attributes, referred to as ``features,''
that are measured on each sample.
Typically $m$ is of the order of a few dozen in small studies, ranging up
to several hundreds for large studies such as the TCGA studies,
while $n$ is of the order of tens of thousands.
There are 20,000 or so genes in the human body, and in whole genome
studies, and the expression level of each gene is measured by at least
one ``probe,'' and sometimes more than one.
The ``raw'' expression level of a gene corresponds to the amount of 
messenger RNA (mRNA) that is produced, and is therefore a nonnegative number.
However, the raw value is often transformed by taking the logarithm
after dividing by a reference value, subtracting a median value, dividing
by a scaling constant, and the like.
As a result the numbers that are reported as gene expression levels
can sometimes be negative numbers.
Therefore it is best to think of gene expression levels as real numbers.
Other features that are measured include micro-RNA (miRNA) levels, 
methylation levels, and copy number variations, all of which can
be thought of as real-valued.
There are also binary features such as the presence or absence of a mutation
in a specific gene.
In addition to these molecular attributes, there are also ``labels''
associated with each tumor.
Let $y_i$ denote the label of tumor $i$, and note that the label depends
only on the sample index $i$ and not the feature index $j$.
Typical real-valued labels include the time of overall survival after
surgery, time to tumor recurrence, or the lethality of a drug on a cancer
cell line.
Typical binary labels include whether a patient had metastasis (cancer
spreading beyond the original site).
In addition, it is also possible for labels to be ordinal variables,
such as ``poor responder,'' ``medium responder,'' and ``good responder.''
Often these ordinal labels are merely quantized version of some other
real-valued attributes.
For instance, the previous example corresponds to a three-level
quantization of the time to tumor recurrence.
In general the labels refer to {\it clinical outcomes}, as in all
of the above examples.
Usually each sample has multiple labels associated with it.
However, in applications, the labels are treated one at a time, so
it is assumed that there is only label for each sample, with $y_i$
denoting the label of the $i$-th sample.
Moreover, for simplicity, it is assumed that the labels are either
real-valued or binary.

Thus the measurement set can be thought of an $m \times n$ matrix
$X = [ x_{ij} ]$, where $x_{ij}$ is the value of feature $j$ in sample $i$.
The row vector $x^i$, denoting the $i$-th row of the matrix $X$, is called
the feature vector associated with sample $i$.
Similarly the column vector $x_j$ denotes the variation of the $j$-th
feature across all $m$ samples.
Throughout this paper, it is assumed that $X \in \R^{m \times n}$, that is,
that each measurement is a real number.
Binary measurements such as the presence or absence of mutations are
usually handled by partitioning the data into two groups, namely those
where the binary measure is zero, and where it is one.
The label $y_i$ is either bipolar (belongs to $\bp$) or is a real number.
Taking the range of two-valued labels $y_i$ as $\bp$ instead of $\bi$
simplifies some of the formulas in the sequel.
If $y_i$ is bipoar the associated problem is called ``classification''
whereas if $y_i$ is real the associated problem is called ``regression.''
In either case, the objective is to find a function $f: \R^n \ap \R$ or
$f: \R^n \ap \bp$ such that $y_i$ is well-approximated by $f(x^i)$.

\section{Regression Methods}\label{sec:5}


The focus in this section is on the case where the label $y_i$ is a real
number.
Therefore the objective is to find a function $f : \R^n \ap \R$ such that
$f(x^i)$ is a good approximation of $y_i$ for all $i$.
%
A typical application in cancer biology would be 
the prediction of the time for a tumor to recur after surgery.
The data would consist of expression levels of tens of thousands
of genes on around a hundred or so tumors, together with the time
for the tumor to recur for each patient.
The objective is to identify a small number of genes whose
expression values would lead to a reliable prediction of the recurrence time.
Cancer is a complex, multi-genic disease, and identifying a small set
of genes that appear to be highly predictive in a particular form of cancer
would be very useful.
Explaining {\it why\/} these genes are the key genes would require
constructing gene regulatory networks.
While this problem is also amenable to treatment using statistical methods,
it is beyond the scope of the present paper.
Towards the end of this section, the tumor recurrence problem is studied using
a new regression method.

Throughout this section, attention is focused on
{\it linear regressors}, with $f(x) = x w - \th$ where
$w \in \R^n$ is a weight vector and $\th \in \R$ is a threshold or bias.
There are several reasons for restricting attention to linear regressors.
From a mathematical standpoint, linear regressors
are by far the most widely studied
and the best understood class of regressors.
From a biological standpoint, it makes sense to suppose that
the measured outcome is a weighted linear combination of each feature,
with perhaps some offset term.
If one were to use higher-order polynomials for example, then
biologists would rightly object that taking the product of two features
(say two gene expression values) is unrealistic most of the time.\footnote{There
are situations such as transcription factor genes regulating other genes, where
taking such a product would be realistic.
But such situations are relatively rare.}
Other possibilities include pre-processing each feature
$x_{ij}$ through a function such as $x \mapsto e^x/(1+e^x)$,
but this is still linear regression in terms of the processed values.
As explained earlier, often the measured feature values $x_{ij}$ are
themselves processed values of the corresponding ``raw'' measurements.

In traditional least-squares regression,
the objective is to choose a weight vector $w \in \R^n$
and a threshold $\th$ so as to minimize the least squared error
\be\label{eq:51}
J_{{\rm LS}} := \sum_{i=1}^m ( x^i w - \th - y_i )^2 .
\ee
This method goes back to Legendre and Gauss, and is the staple
of researchers everywhere.
Let $\eb$ denote a column vector of all ones,
with the subscript denoting the dimension.
Then
\bd
J_{{\rm LS}} = \nm X w - \th \eb_m - y \nm_2^2 
= \nm \bar{X} \bar{w} - y \nm_2^2 ,
\ed
where
\bd
\bar{X} = [ \ba{cc} X & - \eb_m \ea ] \in \R^{m \times (n+1)} ,
\bar{w} = \left[ \ba{c} w \\ \th \ea \right] \in \R^{n+1} .
\ed
If the matrix $\bar{X}$ has full column rank of $n+1$, then it is easy
to see that the unique optimal choice $\bar{w}^*$ is given by
\bd
\bar{w}_{{\rm LS}}^* = ( \bar{X}^t \bar{X} )^{-1} \bar{X}^t y =
\left[ \ba{cc}
X^t X & - X^t \eb_m  \\ - \eb_m^t X & m \ea \right]^{-1}
\left[ \ba{c} X^t \\ \eb_m^t \ea \right] y .
\ed
In the present context, the fact that $m < n$ ensures that the matrix
$X$ has rank less than $n$, whence the matrix $\bar{X}$ has rank
less than $n+1$.
As a result, the standard least squares regression problem does not
have a unique solution.
Therefore one attempts to minimize the least-squares error while imposing
various constraints (or penalties) on the weight vector $w$.\footnote{Note
that no penalty is imposed on the threshold $\th$.}
Different constraints lead to different problem formulations.
An excellent and very detailed treatment of the various topics of this
bsection can be found in \cite[Chapter 3]{Hastie-Tib-Fri11}.

Suppose we minimize the least-squared error objective function subject
to an $\ell_2$-norm constraint on $w$.
This approach to finding a unique set of weights is known as ``ridge
regression'' and is usually credited to \cite{Ridge}.
However, it would perhaps be fairer to credit the Russian mathematician
A.\ N.\ Tikhonov \cite{Tikhonov43}.
In ridge regression the problem is reformulated as
\bd
\min \sum_{i=1}^m ( x^i w - \th - y_i )^2 
\st \nm w \nm_2 \leq t ,
\ed
where $t$ is some prespecified bound.
In the associated Lagrangian formulation, the problem becomes one of
minimizing the objective function
\be\label{eq:52}
J_{{\rm ridge}} := \sum_{i=1}^m ( x^i w - \th - y_i )^2
+ \l \nm w \nm_2^2 ,
\ee
where $\l$ is the Lagrange multiplier.
Because of the additional term,
the $(1,1)$-block of the Hessian of $J_{{\rm ridge}}$,
which is the Hessian of $J_{{\rm ridge}}$ with respect to $w$, now equals
$\l I_n + X^t X$ which is positive definite even when $m < n$.
Therefore the overall Hessian matrix is positive definite under
a mild technical condition, and the problem has a unique solution
for every value of the Lagrange parameter $\l$.
However, the major disadvantage of ridge regression
is that, in general, {\it every component\/}
of the optimal weight vector $w_{{\rm ridge}}$ is nonzero.
In the context of biological applications, this means that the regression
function makes use of {\it every\/} feature $x_j$, which is in general
undesirable.

Another possibility is to choose a solution $w$ that has the 
fewest number of nonzero components, that is, a regressor that
uses the fewest number of features.
Define
\bd
\nm w \nm_p := \left( \sum_{i=1}^n | w_i |^p \right)^{1/p}
\ed
If $p \geq 1$, this is the familiar $\ell_p$-norm.
If $p < 1$, this quantity is no longer a norm, as the function
$w \mapsto \nm w \nm_p$ is no longer convex.
However, as $p \downarrow 0$, the quantity $\nm w \nm_p$
approaches the number of nonzero components of $w$.
For this reason it is common to refer to the ``$\ell_0$-norm'' even
though $\nm \cdot \nm_0$ is not a norm at all.
Moreover, it is known \cite{Natarajan95} that the problem of finding
a $\bar{w}$ such that $\nm w \nm_0$ is minimized is NP-hard.

A very general formulation of the regression problem is to minimize
\be\label{eq:53a}
J_{{\rm M}} := \sum_{i=1}^m ( x^i w - \th - y_i )^2 + \Rcal(w),
\ee
where $\Rcal : \R^n \ap \R_+$ is a norm known as the ``regularizer.''
This problem is analyzed at a very high level of generality
in \cite{NRWY12}, where the least-squares error term is replaced by
an arbitrary convex ``loss'' function.
In the interests of simplicity, we do not discuss the results of
\cite{NRWY12} in their full generality, and restrict the discussion
to least-squares regression alone.

In \cite{Tibshirani-Lasso} it is proposed to
minimize the least-squared error objective function
subject to an $\ell_1$-norm constraint on the weight vector $w$.
In Lagrangian formulation, the problem is to minimize
\be\label{eq:53}
J_{{\rm lasso}} := \sum_{i=1}^m ( x^i w - \th - y_i )^2
+ \l \nm w \nm_1 ,
\ee
where $\l$ is the Lagrange multiplier.
The acronym ``lasso'' is coined in \cite{Tibshirani-Lasso}, and stands for
``least absolute shrinkage and selection operator''.
The lasso penalty can be rationalized by observing that $\nm \cdot \nm_1$
is the convex relaxation of the ``$\ell_0$-norm.''
The behavior of the solution to the lasso algorithm depends on the
choice of the upper bound $t$.
A detailed analysis of the Lagrangian formulation (\ref{eq:53}) and its
dual problem is carried out in \cite{Osborne-Presnell-Turlach00}.
It is shown there that, if the Lagrange multiplier $\l$ in (\ref{eq:53})
is sufficiently large, say $\l > \l_{{\rm max}}$, then the only solution
to the lasso minimization problem is $w = 0$.
Moreover, the threshold $\l_{{\rm max}}$ is not easy to estimate {\it a priori}.
An optimal solution is defined to be ``regular'' in
\cite[Definition 3.3]{Osborne-Presnell-Turlach00}
if it satisfies some technical conditions.
In every problem there is at least one regular solution.
Moreover, every regular optimal  weight vector has at most $m$
nonzero entries; see \cite[Theorem 3.5]{Osborne-Presnell-Turlach00}.

In many applications, some of the columns of the matrix $X$ are highly
correlated.
For instance, if the indices $j$ and $k$ correspond to two genes that
are in the same biological pathway, then their expression levels
would vary in tandem across all samples.
Therefore the column vectors $x_j$ and $x_k$ would be highly correlated.
In such a case, ridge regression tends to
assign nearly equal weights to each.
At the other extreme, lasso tends to choose just one amongst the many
correlated columns and to discard the rest;
which one gets chosen is often a function
of the ``noise'' in the measurements.
In biological data sets, it is reasonable to expect that expression
levels of genes that are in a common pathway are highly correlated.
In such a situation, it is undesirable to choose just one amongst
these genes and to discard the rest;
it is also undesirable to choose all of them, as that would lead to too
many features being chosen.
It would be desirable to choose more than one, but not all, of
the correlated columns.
This is achieved by the so-called ``elastic net'' algorithm, 
introduced in \cite{Zou-Hastie05},
which is a variation of the lasso algorithm.
In this algorithm, the penalty aims to constrain, not the $\ell_1$-norm of the
weight $w$, but a weighted sum of its $\ell_1$-norm and $\ell_2$-norm
squared.
The problem formulation in this case, in Lagrangian form,
is to choose $w$ so as to minimize
\be\label{eq:55}
J_{{\rm EN}} :=  \sum_{i=1}^n ( x^i w - \th - y_i )^2
+ \l [ \mu \nm w \nm_2^2 + (1 - \mu) \nm w \nm_1 ] ,
\ee
where $\mu \in (0,1)$.
Note that if $\mu = 0$, then the elastic net algorithm becomes
the lasso, whereas with $\mu = 1$, the elastic net algorithm becomes 
ridge regression.
Thus the elastic net algorithm provides a bridge between the two.
Note that the penalty term in the elastic net algorithm is {\it not\/}
a norm, due to the presence of the squared term;
hence the elastic net algorithm is not covered by the very thorough
analysis in \cite{NRWY12}.
A useful property of the elastic net algorithm is brought out in
\cite[Theorem 1]{Zou-Hastie05}.

\begin{theorem}\label{thm:EN}
Assume that $y,X,\l$ are fixed, and let $\bar{w}$ denote the 
corresponding minimizer of (\ref{eq:55}).
Assume without loss of generality that $y$ is centered, that is,
$y^t \eb_m = 0$, and that the columns of $X$ are normalized such that
$\nm x_j \nm_2 = 1$ for all $j$.
Let $j,k$ be two indices between $1$ and $n$, and
suppose that $x_j^t x_k \geq 0$.
Then
\be\label{eq:55a}
| w_j - w_k | \leq \frac{ \nm y \nm_1 }{ \l \mu } \sqrt{ 2 ( 1 - x_j^t x_k ) } .
\ee
\end{theorem}
Since one can always ensure that $x_j^t x_k^t \geq 0$ by replacing $x_k$
by $- x_k$ if necessary, (\ref{eq:55a}) states that if the columns
$x_j$ and $x_k$ are highly correlated, then the corresponding
coefficients in the regressor are nearly equal.
Unlike in the lasso algorithm, there do not seem to be many results
on the number of nonzero weights that are chosen by the elastic net algorithm.
It can and often does happen that the number of features chosen
is larger than $m$, the number of samples.
However, as explained above, this is often seen as a desirable feature
when the columns of the matrix $X$ are highly correlated, as they often
are in biology data sets.

Next we discuss several versions of the problem formulation in
(\ref{eq:53a}) corresponding to diverse choices of the penalty norm $\Rcal$,
culminating in some open problems that are relevant
to biological applications.
The ``pure'' lasso algorithm tries to choose as few distinct features
as possible in the regressor.
However, it may be worthwhile to partition
the set of features $\N =\{ 1 , \ldots , n \}$
into $g$ groups $G_1 , \ldots , G_g$, and then choose a regressor
that selects elements from as few distinct groups as possible,
without worrying about the number of features chosen.
This is achieved by the so-called group lasso algorithm introduced
in \cite{Bakin99} and \cite{Lin-Zhang06}.
Let $n_l := | G_l |$ for $l = 1 , \ldots , g$.
In the grouped lasso algorithm, the objective function is
\be\label{eq:54}
J_{{\rm GL}} = \sum_{i=1}^m ( x^i w - \th - y_i )^2
+ \l \sum_{l=1}^g \sqrt{n_l} \nm w_{G_l} \nm_2 ,
\ee
where $w_{G_l} \in \R^n$ is determined from $w$ by setting $w_j = 0$
for all $j \not \in G_l$.
It is clear that, depending on the relative sizes of the various groups,
one weight vector can have more nonzero components than another,
and yet the number of distinct groups to which these nonzero components
belong can be smaller.
In the limiting case, if the number of groups is taken as $n$ and
each group is taken to consist of a singleton set, then the grouped
lasso reduces to the standard lasso algorithm.
A further variation is the so-called sparse group lasso algorithm
introduced in \cite{FHT10,SFHT12}, where the objective is simultaneously
to choose features from as few distinct groups as possible, and within
the chosen groups, choose as few features as possible.
The objective function in the sparse group lasso (SGL) algorithm is
\be\label{eq:54a}
J_{{\rm SGL}} = \sum_{i=1}^m ( x^i w - \th - y_i )^2
+ \l \sum_{l=1}^g [ ( 1 - \mu ) \nm w_{G_l} \nm_1 
+ \mu \nm w_{G_l} \nm_2 ] ,
\ee
where as always $\mu \in [0,1]$.

The above formulations of the GL and SGL norms
based on the assumption that the various groups do not overlap.
However, in some biological applications it makes sense to permit
overlapping group decompositions.
Specifically, at a first-level of approximation a gene regulatory network
(GRN) can be modelled as a directed acyclic graph (DAG), wherein the
root nodes can be interpreted as master regulator genes, and directed
paths can be interpreted as biological pathways.
In such a case, one seeks to explain the available data, not by
choosing the fewest number of {\it genes}, but rather by the fewest
number of {\it pathways}.
To illustrate, consider the baby example shown in Figure \ref{fig:1},
where genes 1 and 2 are master regulators, while genes 3 through 7 are
regulated genes.
Some are regulated directly by a master regulator gene, while others
are indirectly regulated.
In Figure \ref{fig:1}(i), there are four pathways, namely
\bd
G_1 = \{ 1 , 2 , 4 \} , G_2 = \{ 1, 2, 5 \} ,
G_3 = \{ 1, 3, 6 \} , G_4 = \{ 1, 3, 7 \} ,
\ed
whereas in Figure \ref{fig:1}(ii) there are also four pathways, namely
\bd
G_1 = \{ 1 , 2 , 4 \} , G_2 = \{ 1, 2, 5 \} ,
G_3 = \{ 1, 3, 5 \} , G_4 = \{ 1, 3, 6 \} .
\ed
Ideally, we would like to choose a set of features that intersect
with as few pathways as possible.
We will return to this example after presenting available theories
for sparse regression with overlapping groups.

\bfig

\btab{cc}

\begin{minipage}{60mm}
\bc
\btp[line width = 2pt]


\draw (0,0) node [ minimum size = 8 mm , draw , circle ] {1} ;
\draw (-1.5,-1) node [ minimum size = 8 mm , draw , circle ] {2} ;
\draw (1.5,-1) node [ minimum size = 8 mm , draw , circle ] {3} ;
\draw (-2.25,-2) node [ minimum size = 8 mm , draw , circle ] {4} ;
\draw (-0.75,-2) node [ minimum size = 8 mm , draw , circle ] {5} ;
\draw (0.75,-2) node [ minimum size = 8 mm , draw , circle ] {6} ;
\draw (2.25,-2) node [ minimum size = 8 mm , draw , circle ] {7} ;


\draw [->] (-0.28,-0.28) -- (-1.22,-0.72) ;
\draw [->] (-1.78,-1.28) -- (-1.97,-1.72) ;
\draw [->] (-1.22,-1.28) -- (-0.97,-1.72) ;
\draw [->] (0.28,-0.28) -- (1.22,-0.72) ;
\draw [->] (1.78,-1.28) -- (1.97,-1.72) ;
\draw [->] (1.22,-1.28) -- (0.97,-1.72) ;

\etp
\ec
\end{minipage}

&

\begin{minipage}{60mm}
\bc
\btp[line width = 2pt]


\draw (0,0) node [ minimum size = 8 mm , draw , circle ] {1} ;
\draw (-1,-1) node [ minimum size = 8 mm , draw , circle ] {2} ;
\draw (1,-1) node [ minimum size = 8 mm , draw , circle ] {3} ;
\draw (-2,-2) node [ minimum size = 8 mm , draw , circle ] {4} ;
\draw (0,-2) node [ minimum size = 8 mm , draw , circle ] {5} ;
\draw (2,-2) node [ minimum size = 8 mm , draw , circle ] {6} ;


\draw [->] (-0.28,-0.28) -- (-0.72,-0.72) ;
\draw [->] (-1.28,-1.28) -- (-1.72,-1.72) ;
\draw [->] (-0.72,-1.28) -- (-0.28,-1.72) ;
\draw [->] (0.28,-0.28) -- (0.72,-0.72) ;
\draw [->] (1.28,-1.28) -- (1.72,-1.72) ;
\draw [->] (0.72,-1.28) -- (0.28,-1.72) ;

\etp
\ec
\end{minipage}

\etab

\caption{
Two regulatory networks.
(i) A network without overlapping groups.
(ii) A network with overlapping groups.
}

\label{fig:1}

\efig

To date, various versions of group or sparse group lasso
with overlapping groups have been proposed.
As before, let $G_1 , \ldots , G_g$ be subsets of $\N = \{ 1 , \ldots , n \}$,
but now {\it without\/} the assumption that the groups are pairwise disjoint.
The penalty-augmented optimization problems are the same as in
(\ref{eq:54}) and (\ref{eq:54a}) respectively; however, the objective
functions are now referred to as $J_{{\rm GLO}}$ and $J_{{\rm SGLO}}$
to suggest (sparse) group lasso with overlap.
For the case of overlapping groups, the theory developed in \cite{NRWY12}
continues to apply so long as the penalty terms in (\ref{eq:54})
and (\ref{eq:54a}) respectively are ``decomposable.''
The most general results available to date address the case where
the groups are ``tree-structured,'' that is,
\be\label{eq:56}
G_i \cap G_j \neq \es \imp \{ G_i \seq G_j \mbox{ or } G_j \seq G_i \} .
\ee
See for example \cite{OJV-Over-GL-11,Jenetton-et-al11}.

Now, if we examine the groups associated with the network in Figure
\ref{fig:1}(i), it is obvious that (\ref{eq:56}) is not satisfied.
However, there is a slight modification that would permit
(\ref{eq:56}) to hold, namely to drop the root node and retain only
the successors.
Thus the various groups are
\bd
G_1 = \{ 4 \} , G_2 = \{ 5 \} , G_3 = \{ 6 \} , G_4 = \{ 7 \} ,
\ed
\bd
G_5 = \{ 2, 4 \} , G_6 = \{ 2, 5 \} , G_7 = \{ 3, 6 \} , G_8 = \{ 3, 7 \} .
\ed
However, there is no way of modifying the groups so as to
ensure that (\ref{eq:56}) holds for
the network in Figure \ref{fig:1}(ii).
The reason is easy to see.
The ``tree structure'' assumption (\ref{eq:56}) implies that
there is only one path between every pair of nodes.
But this is clearly not true in Figure \ref{fig:1}(ii), because
there are two distinct paths from node 1 to node 5.
Moreover, a little thought would reveal that that the assumption
of tree-structured groups does not really permit truly overlapping
groups.
In particular, if (\ref{eq:56}) holds, then the collection of sets
$\{ G_1 , \ldots , G_g \}$ can be expressed as a union of chains
in the form
\bd
G_{11} \seq \ldots \seq G_{1g_1} , \ldots , 
G_{s1} \seq \ldots \seq G_{sg_s} ,
\ed
where the ``maximal'' sets $G_{ig_i}$ are pairwise disjoint
and together span the total feature set $\N = \{ 1 , \ldots , n \}$.
Now, in a biological network, it makes no sense to impose a condition
that there must be only path between every pair of nodes.
Therefore the problem of defining a decomposable norm penalty
for inducing other types of sparsity besides tree-structure,
especially the types of sparsity that are consistent with biology, 
is still open.

We conclude this section with a new algorithm and its application to
sparse regression.
This represents joint work with Mehmet Eren Ahsen
and will be presented in more complete form elsewhere.
A special case of SGL
is obtained by choosing just one group, which perforce has to equal $\N$,
so that
\be\label{eq:54b}
J_{{\rm MEN}} = \sum_{i=1}^m ( x^i w - \th - y_i )^2
+ \l [ ( 1 - \mu ) \nm w \nm_1 + \mu \nm w \nm_2 ] .
\ee
Of course, since the entire index set $\N$ is chosen as one group,
there is nothing ``sparse'' about it.
Note that the only difference between (\ref{eq:54b}) and (\ref{eq:55})
is that the $\ell_2$-norm is {\it not\/} squared in the former.
For this reason, the above approach is called the `` modified elastic net''
or MEN algorithm.
Unlike in EN, the penalty (or constraint) term in MEN is a norm,
being a convex combination of the $\ell_1$- and $\ell_2$-norms.
In several examples, the MEN algorithm appears to combine the accuracy
of EN with the sparsity of lasso.
It is relatively easy to prove an analog of Theorem \ref{thm:EN}
for the MEN algorithm.
That is, unlike in lasso but as in EN,
MEN assigns nearly equal weights to highly correlated features.
But further theoretical analysis remains to be carried out.

The MEN algorithm was applied to the TCGA ovarian cancer data
(\cite{TCGA-Ovarian}) to predict the time to tumor recurrence.
Specifically, both times to tumor recurrence as well as
expression levels for 12,042 genes are available for 283 patients.
Out of these, 40 patients whose tumors recurred before 210 days 
or after 1,095 days were excluded from the study as being
``extreme'' cases.
The remaining 243 samples were analyzed using MEN with recursive
feature elimination.
The results are shown in Figure \ref{fig:tumor}.
The number of features and the average percentage error in absolute value
are shown in Table \ref{table:tumor}.

\begin{table}[h]
\bc
\btab{|l|r|r|}
\hline
{\bf Algorithm} & {\bf No.\ of} & {\bf Average} \\
& {\bf Features} & {\bf Perc.\ Error} \\
\hline
LASSO & 43 & 16.14\% \\
Elastic Net & 60 & 14.35\& \\
MEN & 42 & 14.91\% \\
\hline
\etab
\ec

\caption{Comparison of three algorithms on TCGA ovarian cancer data
on time to tumor recurrence, with extreme cases excluded.}

\label{table:tumor}

\end{table}

\bfig

\bc
\btab{c}

\begin{minipage}{80mm}
\includegraphics[width=80mm]{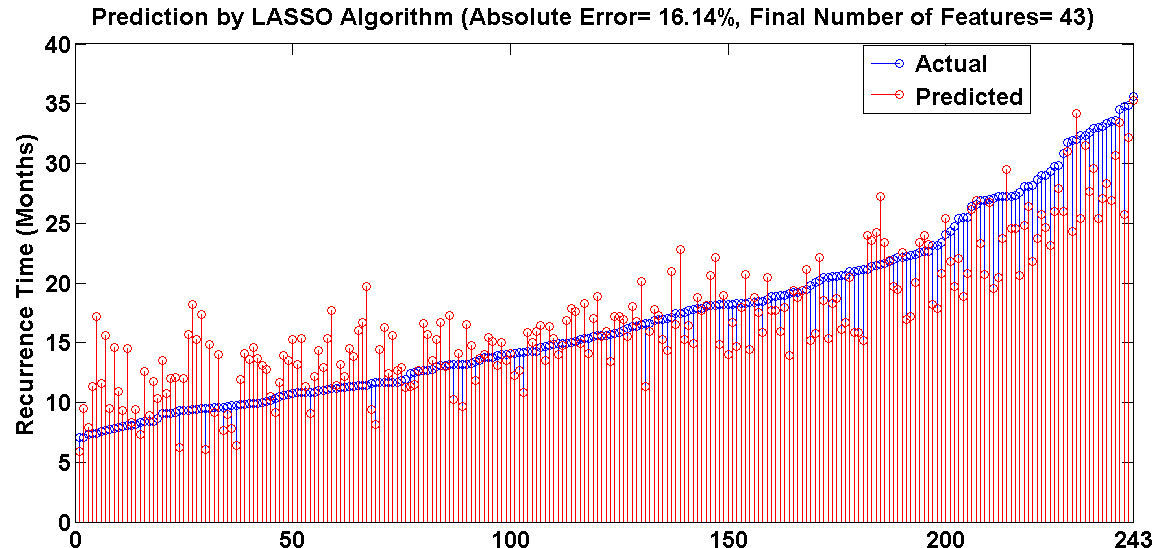}
\end{minipage}

\\

\begin{minipage}{80mm}
\includegraphics[width=80mm]{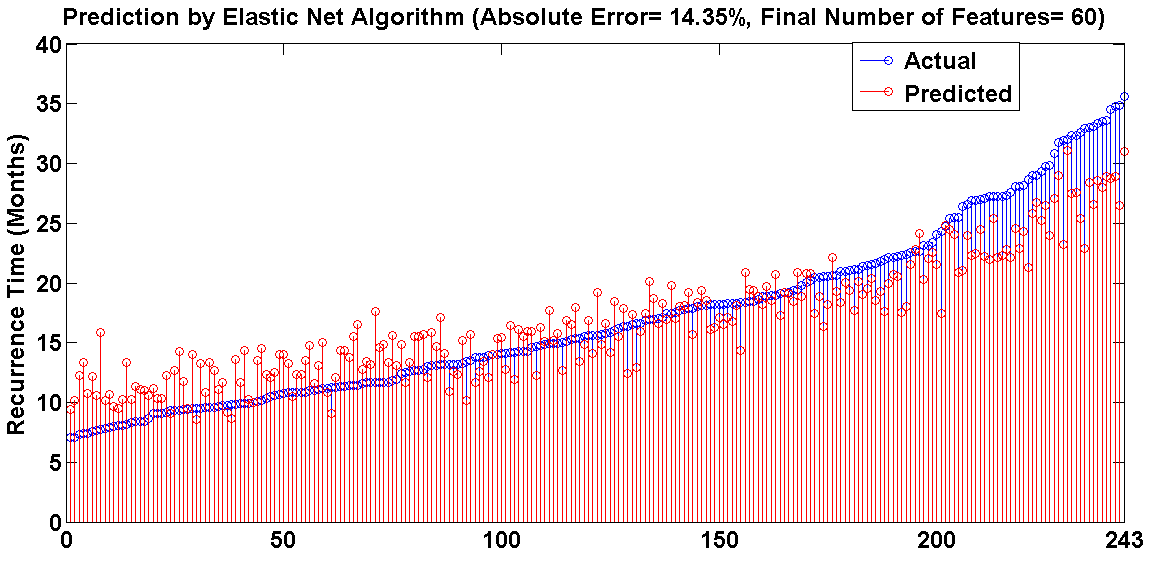}
\end{minipage}

\\

\begin{minipage}{80mm}
\includegraphics[width=80mm]{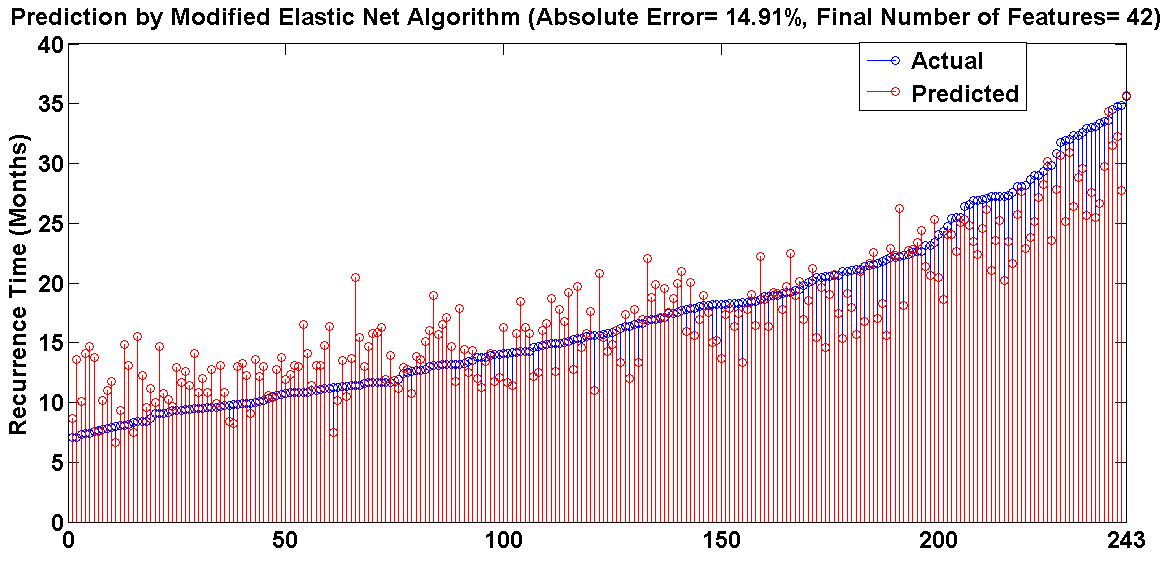}
\end{minipage}

\etab
\ec

\caption{Predicted vs.\ actual times to tumor recurrence in 243
ovarian cancer patients.
The results for lasso are at the top, those for the elastic net are
in the middle and those for the modified elastic net algorithm are at
the bottom.
}

\label{fig:tumor}

\efig

\section{Compressed Sensing}\label{sec:6}

In recent years, there have been several results that are grouped
under the general heading of ``compressed sensing'' or
``compressive sensing''.
Both expressions are in use, but ``compressed sensing'' is used in this paper.
The problem can be roughly stated as follows:
Suppose $x \in \R^n$ is an unknown vector but with known structure;
is it possible to determine $x$ either exactly or approximately,
by taking $m \ll n$ linear measurements of $x$?
The area of research that goes under this broad heading grew spectacularly
during the first decade of the new millennium.\footnote{In \cite{DDEK12}
it is suggested a precursor of compressed sensing can be found in
a paper that dates back to 1795!}
As summarized in the introduction of the paper \cite{Donoho06a},
the impetus for recent work in this area was the desire to find
algorithms for data compression that are ``universal'' in the sense of
being nonadaptive (i.e., do not depend on the data).
In the original papers in this area, the results and proofs were a mixture
of sampling, signal transformation (time domain to frequency domain
and vice versa), randomness etc.
However, as time went on, the essential ingredients of the approach
were identified, thus leading to a very streamlined theory that
clearly transcends its original application domains of image and signal
processing.

The motivation for discussing compressed sensing theory in the present
paper is the following:
Whether it is in compressed sensing or in computational biology,
one searches for a relatively simple explanation of the observations.
Therefore it may {\it potentially\/} be possible to borrow some of
the basic ideas from compressed sensing
theory and adapt them to problems in cancer biology.
Compressed sensing theory {\it as it currently stands\/}
cannot directly be applied to the analysis of biological data sets,
because the fundamental assumption in compressed sensing
theory is that {\it one is able to choose\/} the so-called measurement
matrix, called $A$ throughout this paper.
Note that in statistics, the matrix $A$ is often referred to as the
``design'' matrix.
However, in biological applications this matrix is often fixed,
and one does not have the freedom to choose, i.e.\ to ``design'' it.
However, in biological (and other) applications, the measurement matrix
is given, and one does not have the freedom to change it.
Nevertheless, the developments in this area are too important to be ignored
by computational biologists.
The hope is that, by understanding the core arguments of compressed
sensing theory and building on them,
it will be possible for the computational biology community
to develop a similarly successful theory for their application domain.
Therefore an introductory treatment of compressed sensing is included here.

The major developments in this area are generally associated with
the names of Cand\`{e}s, Donoho, Romberg, and Tao, though several
other researchers have also made important contributions.
See \cite{Donoho06a} for one of the earliest comprehensive papers,
as well
\cite{Donoho06a,Donoho06b,Candes08,Candes-Tao05,Candes-Tao07,Candes-Plan09,Romberg09,Cohen-Dahmen-Devore09}.
The survey paper \cite{DDEK12} and a recent paper
\cite{NRWY12} contain a wealth of bibliographic
references that can be followed up by interested readers.

We begin by introducing some notation.
Suppose $m,n,k$ are given integers, with $n \geq 2k$.
For convenience, we denote the set $\{ 1 , \ldots , n \}$ by $\N$ throughout.
For a given vector $x \in \R^n$, let $\supp(x)$ denote its support,
that is, $\supp(x) = \{ i: x_i \neq 0 \}$.
Let $\SI_k = \{ x \in \R^n : | \supp(x) | \leq k \}$.
Thus $\SI_k$ denotes the set of ``$k$-sparse'' vectors in $\R^n$, or in
other words, the set of $n$-dimensional vectors that have $k$ or fewer
nonzero components.
For each vector $x \in \R^n$, integer $k < n$, and norm $\nm \cdot \nm$
on $\R^n$, the symbol $\s_k(x,\nm \cdot \nm)$ denotes the distance from
$x$ to $\SI_k$, that is,
\bd
\s_k(x,\nm \cdot \nm) = \inf \{ \nm x - z \nm : z \in \SI_k \} .
\ed
The quantity $\s_k(x,\nm \cdot \nm)$ is called the ``sparsity measure''
of the vector $x$ of order $k$ with respect to the norm $\nm \cdot \nm$.
It is obvious that $\s_k(x,\nm \cdot \nm)$ depends on the underlying norm.
However, if  $\nm \cdot \nm$ is one of the $\ell_p$-norms, then it is
easy to compute $\s_k(x,\nm \cdot \nm)$.
Specifically, given $k$, let $\L_0$ denote the index set corresponding
to the $k$-largest components of $x$ in magnitude, and let $\xloc$
denote the vector that results by replacing the components of $x$ 
in the set $\L_0$ by zeros.
(It is convenient to think of $\xlc$ as an element of $\R^n$ rather
than an element of $\R^{n-k}$.)
Then, whenever $p \in [1, \infty]$, it is easy to see that
\bd
\s_k(x,\nm \cdot \nm_p) = \nm \xloc \nm_p .
\ed

Next, the so-called `` restricted isometry property'' (RIP) is introduced.
Note that in some cases the RIP can be replaced by a weaker property
known as
the ``null space property'' (\cite{Cohen-Dahmen-Devore09}).
However, the objective of the present paper is {\it not\/} to
present the most general results, but rather to present reasonably general
results that are easy to explain.
So the exposition below is confined to the RIP.

\begin{definition}\label{def:61}
Suppose $A \in \R^{m \times n}$.
We say that $A$ satisfies the {\bf Restricted Isometry Property (RIP)}
of order $k$ with constant $\d_k$ if
\be\label{eq:61}
(1 - \d_k) \nm u \nm_2^2 \leq \langle u , A u \rangle \leq
(1 + \d_k) \nm u \nm_2^2 , \fa u \in \SI_k .
\ee
\end{definition}

So the matrix $A$ has the RIP of order $k$ with constant $1 - \d_k$
if the following property holds:
For every choice of $k$ or fewer columns of $A$ (say the columns in
the set $J \seq \N$ where $|J| \leq k$), the spectrum of the symmetric
matrix $A_J^t A_J$ lies in the interval $[1 - \d_k , 1 + \d_k ]$,
where $A_J \in \R^{m \times |J|}$ denotes
the submatrix of $A$ consisting of all rows and the columns corresponding
to the indices in $J$.

If integers $n,k$ are specified, the integer $m$ has to be sufficiently
large in order for the matrix $A$ to satisfy the RIP.

\begin{theorem}\label{thm:61}
(\cite[Theorem 1.4]{DDEK12})
Suppose $A \in \R^{m \times n}$ satisfies the RIP or order $2k$
with constant $\d_{2k} \in (0,1/2]$.
Then
\be\label{eq:62}
m \geq c k \log (n/k) = ck ( \log n - \log k) ,
\ee
where
\bd
c = \frac{1}{ 2 \log ( \sqrt{24} +1 ) } \approx0.28 .
\ed
\end{theorem}

Next we state some of the main known results in compressed sensing.
The theorem statement below corresponds to \cite[Theorem 1.2]{Candes08}
and \cite[Theorem 1.9]{DDEK12}.
It can be compared with \cite[Theorem 1.4]{Candes-Plan09}).

\begin{theorem}\label{thm:63}
Suppose $A \in \R^{m \times n}$ satisfies the RIP of order $\d_{2k}$
with constant $\d_{2k} < \sqrt{2} - 1$, and that $y = A x + \eta$
for some $x \in \R^n$ and $\eta \in \R^m$ with
$\nm \eta \nm_2 \leq \e$.
Let $\B_\e(y) := \{ z \in \R^n : \nm y - Az \nm_2 \leq \e \}$, and define
\be\label{eq:618}
\xh = \argmin_{z \in \B_\e(y)} \nm z \nm_1 .
\ee
Then
\be\label{eq:619}
\nm \xh - x \nm_2 \leq C_0 \frac{ \s_k(x, \nm \cdot \nm_1) }{ \sqrt{k} } 
+ C_2 \e ,
\ee
where
\be\label{eq:616}
C_0
= 2 \frac{ 1 + ( \sqrt{2} - 1 ) \d_{2k} } { 1 - ( \sqrt{2} + 1 ) \d_{2k} } ,
C_2
= \frac{ 4 \sqrt{ 1 + \d_{2k} } } { 1 - ( \sqrt{2} + 1 ) \d_{2k} } .
\ee
\end{theorem}

The formula for $C_2$ is written slightly differently from that in
\cite[Theorem 1.9]{DDEK12} but is equivalent to it.

\begin{corollary}\label{corr:62}
Suppose $A \in \R^{m \times n}$ satisfies the RIP of order $\d_{2k}$
with constant $\d_{2k} < \sqrt{2} - 1$, and that $y = A x + \eta$
for some $x \in \SI_k$ and $\eta \in \R^m$ with
$\nm \eta \nm_2 \leq \e$.
Let $\B_\e(y) := \{ z \in \R^n : \nm y - Az \nm_2 \leq \e \}$, and define
\be\label{eq:618a}
\xh = \argmin_{z \in \B_\e(y)} \nm z \nm_1 .
\ee
Then
\be\label{eq:619a}
\nm \xh - x \nm_2 \leq C_2 \e ,
\ee
where $C_2$ is defined in (\ref{eq:616}).
\end{corollary}

\begin{corollary}\label{corr:61}
Suppose $A \in \R^{m \times n}$ satisfies the RIP of order $\d_{2k}$
with constant $\d_{2k} < \sqrt{2} - 1$, and that $y = A x$ for some
$x \in \SI_k$.
Let $A^{-1}(y) := \{ z \in \R^n : y = A z \}$, and define
\be\label{eq:617}
\xh = \argmin_{z \in A^{-1}(y)} \nm z \nm_1 .
\ee
Then $\xh = x$.
\end{corollary}

Both corollaries follow readily from the bound (\ref{eq:619}).
Note that if $x \in \SI_k$ then $\s_k(x, \nm \cdot \nm_1) = 0$.
Thus (\ref{eq:619}) implies that $\nm \xh - x \nm_2 \leq C_2 \e$
if there is measurement error, and
$\nm \xh - x \nm_2 = 0$, i.e.\ that
$\xh = x$, if there no measurement error.

Corollary \ref{corr:62} is a simplified version of a corresponding
result in \cite{Candes-Plan09} and is referred to as the ``near ideal''
property of the lasso algorithm.
Suppose that $x \in \SI_k$ so that $x$ is $k$-sparse.
Let $S$ denote the support of $x$, and let $A_S \in \R^{m \times |S|}$
denote the submatrix of $A$ consisting of the columns corresponding
to indices in $S$.
If an ``oracle'' knew not only the size of $S$, but the set $S$ itself,
then the oracle could compute $\xh$ as
\bd
\xh_{{\rm oracle}} = (A_S^T A_S)^{-1} A_S^T y 
= x +  (A_S^T A_S)^{-1} A_S^T \eta . 
\ed
Then the error would be
\bd
\nm \xh_{{\rm oracle}} - x \nm_2 = \nm (A_S^T A_S)^{-1} A_S^T \eta \nm_2
\leq \cons \cdot \e 
\ed
for some appropriate constant.
On the other hand, if $x \in \SI_k$, then $\s_k(x,\nm \cdot \nm_1) = 0$,
and the right side of (\ref{eq:619}) reduces to (\ref{eq:619a}), that is,
\bd
\nm \xh -  x \nm_2 \leq C_2 \e .
\ed
The point therefore is that, if the matrix $A$ satisfies RIP,
and the constant $\d_{2k}$ satisfies the ``compressibility condition''
$\d_{2k} < \sqrt{2} - 1$,
then the mean-squared error of
the solution to the optimization problem (\ref{eq:618a})
is bounded by a fixed (or ``universal'') constant times the
error bound achieved by an ``oracle'' that knows the support of $x$.

The advantage of the above theorem statements, which are taken from
\cite{Candes08,DDEK12}, is that the role of various conditions is clearly
delineated.
For instance, the construction of a matrix $A \in \R^{m \times n}$
that satisfies the RIP is usually achieved by some randomized algorithm.
In \cite[Theorem 1.5]{Candes-Tao05} such a matrix is constructed by
taking the columns of $A$ to be samples of i.i.d.\ Gaussian variables.
In \cite{Achlioptas03}, Bernoulli processes are used to construct
$A$, which has the advantage of ensuring that all elements $a_{ij}$
have just three possible values, namely $0,+1,-1$.
A simple proof that the resulting matrices satisfy the RIP with
high probability is given in \cite{BDDW08}.
Neither of these construction methods is {\it guaranteed\/}
to generate a matrix $A$ that satisfies RIP.
Rather, the resulting matrix $A$ satisfies RIP with some probability,
say $\geq 1 - \g_1$.
The probability $\g_1$ that the randomized method may fail to generate
a suitable $A$ matrix can be bounded using techniques that have nothing
to do with the above theorem.
Similarly, in case the measurement matrix $A$ satisfies the RIP but
the measurement noise $\eta$ is random, then it is
obvious that Theorem \ref{thm:63} holds with probability $\geq 1 - \g_2$,
where $\g_2$ is a bound on the tail probability $\Pr \{ \nm \eta \nm_2
> \e \}$.
Again, the problem of bounding this tail probability has nothing to do
with Theorem \ref{thm:63}.
By combining both estimates, it follows that if the measurement matrix
$A$ is generated through randomization, and if the measurement noise
is also random, then Theorem \ref{thm:63} holds with probability
$\geq 1 - \g_1 - \g_2$.

Observe that the optimization problem (\ref{eq:618a}) is
\bd
\min_z \nm z \nm_1 \st \nm y - Az \nm_2 \leq \e .
\ed
This is essentially the same as the lasso algorithm, except that
the objective function and the constraint are interchanged.
This raises the question as to whether the $\ell_1$-norm can
be replaced by some other norm $\nmP{ \cdot }$
that induces some other form of sparsity, for example group sparsity.
If some other norm is used in place of the $\ell_1$-norm,
does the resulting algorithm display near-ideal behavior, as does lasso?
In other words, is there an analog of Theorem \ref{thm:63} if $\nm \cdot \nm_1$
is replaced by another penalty $\nmP{ \cdot }$?
In joint work with Mehmet Eren Ahsen (\cite{MV-Eren-Near-Ideal14}),
the author has proved a very general theorem to the following effect:
Whenever the penalty norm is ``decomposable'' and the measurement matrix
$A$ satisfies a ``group RIP,'' the corresponding algorithm has near-ideal
behavior provided a ``compressibility condition'' is satisfied.
The result is described in brief.

Let $\G = \{ G_1 , \ldots , G_g\}$ be a partition of
$\N = \{ 1 , \ldots , n \}$.
This implies that the sets $G_i$ are pairwise disjoint.
If $S \seq \{ 1 , \ldots , g \}$, define $G_S := \cup_{i \in S} G_i$.
Let $k$ be some integer such that $k \geq \max_i | G_i |$.
A subset $\L \seq \N$ is said to be {\bf $S$-group $k$-sparse} if
$\L \seq G_S$ and $| G_S | \leq k$, and {\bf group $k$-sparse} if
it is $S$-group $k$-sparse for some set $S \seq \{ 1 , \ldots , g \}$.
The symbol $\GkS \seq 2^\N$ denotes the collection of group $k$-sparse sets.

Suppose $\nmP { \cdot } : \R^n \ap \R_+$ is some norm.
The next definition builds on an earlier definition from \cite{NRWY12}.

\begin{definition}\label{def:decomp}
$\nmP { \cdot }$ is {\bf decomposable} with respect to the partition $\G$ if
the following is true:
Whenever $u, v \in \R^n$ are group $k$-sparse with support sets 
$\L_u \seq G_{S_1}$, $\L_v \seq G_{S_2}$ and the sets $S_1, S_2$
are disjoint, it is true that
\be\label{eq:21}
\nmP { u + v } = \nmP { u } + \nmP { v } .
\ee
The norm $\nmP { \cdot }$ is said to be {\bf regular} if, whenever $\L$ 
is a strict subset of $G_i$ for some index $i$, it is true that
$\nmP { x_\L } \leq \nmP { x_G }$ with equality if and only if
$x_{G \setminus \L} = 0$.
\end{definition}

By adapting the arguments in \cite{NRWY12}, it can be shown that the
group lasso norm used in (\ref{eq:54}), namely
\bd
\nm x \nm_{{\rm GL}} := \sum_{l=1}^g \sqrt{n_l} \nm x_{G_l} \nm_2 ,
\ed
and the sparse group lasso norm used in (\ref{eq:54a})
\bd
\nm x \nm_{{\rm SL}} := \sum_{l=1}^g [ ( 1 - \mu ) \nm x_{G_l} \nm_1 
+ \mu \nm x_{G_l} \nm_2 ] ,
\ed
are both decomposable as well as regular.

Next, the notion of RIP is extended to groups.

\begin{definition}\label{def:GRIP}
A matrix $A \in \R^{m \times n}$ is said to {\bf satisfy the group RIP
of order $k$ with constants $\ru_k,\rb_k$} if
\be\label{eq:521}
0 < \ru_k \leq \min_{ \L \in \GkS} \min_{\supp(z) \seq \L} 
\frac{ \nm Az \nm_2^2 }{ \nm z \nm_2^2} 
\leq \max_{ \L \in \GkS} \max_{\supp(z) \seq \L}
\frac{ \nm Az \nm_2^2 }{ \nm z \nm_2^2} \leq \rb_k .
\ee
\end{definition}

We define $\d_k := ( \rb_k - \ru_k )/2$,
and introduce some constants:
\be\label{eq:523}
c := \min_{\L \in \GkS} \min_{x_\L \neq 0} \frac{ \nmP { \xl } }
{ \nmeu {\xl } } ,
d := \max_{\L \in \GkS} \max_{x_\L \neq 0} \frac{ \nmP { \xl } }
{ \nmeu {\xl } } .
\ee

With these definitions, the following theorem can be proved.

\begin{theorem}\label{thm:531}
Suppose $A \in \R^{m \times n}$ satisfies the group RIP property
of order $2k$ with constants $( \ru_{2k}, \rb_{2k} )$ respectively,
and let $\d_{2k} = ( \rb_{2k} - \ru_{2k} )/2$.
Suppose $x \in \R^n$ and that $y = Ax + \eta$ where $\nmeu { \eta } \leq \e$.
Suppose that the norm $\nmP { \cdot }$ is decomposable, and define
\be\label{eq:523a}
\xh = \argmin_{z \in \R^n} \nmP { z } \st \nmeu { y - Az } \leq \e .
\ee
Suppose that
\be\label{eq:524}
\d_{2k} < \frac { c \ru_k } { d } .
\ee
Then 
\be\label{eq:525}
\nmP { \xh - x} \leq \frac{ 2 }{ 1 - r}
[ 2 \s + ( 1 + r ) \zeta \e ] ,
\ee
and
\be\label{eq:526}
\nmeu { \xh - x} \leq \frac{ 2 }{ c ( 1 - r ) }
[ 2 \s + ( 1 + r ) \zeta \e ] ,
\ee
where
\be\label{eq:527}
r := \frac{ \d_{2k} d }{ c \ru_k } ,
\s := \s_{k,\G}(x , \nmP { \cdot } ) ,
\zeta := \frac{ 2 d \sqrt{ \rb_k } } { \ru_k } ,
\ee
and $c,d$ are defined in (\ref{eq:523}).
\end{theorem}

In the above theorem, (\ref{eq:524}) replaces the ``compressibility'' condition
$\d_{2k} < \sqrt{2} - 1$ of Theorem \ref{thm:63}.
The resemblance of (\ref{eq:526}) to (\ref{eq:619}) is obvious.
Consequently, (\ref{eq:526}) can be readily interpreted as stating
that minimizing the decomposable norm
$\nmP{ \cdot |}$ leads to near-ideal behavior.

\section{Classification Methods}\label{sec:4}

The basic problem of classification can be stated as follows:
Suppose we are given vectors $x^i , i = 1 , \ldots , m$ where each
$x^i \in \R^n$ is viewed as a row vector.
Suppose further that the $m$ vectors are grouped into two sets,
call them $\M_1$ and $\M_2$.
Without loss of generality, renumber the vectors such that
$x_1 , \ldots , x_{m_1}$ constitute the first set and
$x_{m_1+1} , \ldots , x_{m_1 + m_2} = x_m$ constitute the second set.
For future use, define $\M = \{ 1 , \ldots , m \}$, and let
$\M_1 = \{ 1 , \ldots , m_1 \}$, $\M_2 = \{ m_1 + 1 , \ldots , m_1 + m_2 = m \}$
be a partition of $\M$.
Assign a label $y_i = +1$ to the vectors in $\M_1$ and a label $y_i = -1$
to the vectors in $\M_2$.
The objective of (two-class) classification is to find a {\bf discriminant
function} $f : \R^n \ap \R$ such that $f(x^i)$ has the same sign as $y_i$
for all $i$, or equivalently $y_i \cdot \sg(f(x^i)) = 1$ for all $i$.
In the present context, the objective is not merely to find such
a discriminant function, but rather, to find one that uses relatively few
features.

In many ways, classification is an easier problem than regression,
because the sole criterion is that the discriminant function $f(x^i)$
should have the same sign as the label $y_i$  for each $i$.
Thus, if $f$ is a discriminant function, so is $\al f$ for every positive
constant $\al$, and more generally, so is any function $\phi(f)$
whenever $\phi$ is a so-called ``first and third-quadrant function,'' i.e.\
where $\phi(u) > 0$ when $u > 0$ and $\phi(u) < 0$ when $u < 0$.
This gives us great latitude in choosing a discriminant function.

\subsection{The Support Vector Machine}\label{ssec:41}

This section is devoted to the well-known Support Vector Machine (SVM),
first introduced in \cite{Cortes-Vapnik97},
which is amongst the most successful and most widely used
tools in machine learning.

A given set of labelled vectors $\{ (x^i,y_i) , x^i \in \R^n , y_i \in \bp \}$
is said to be {\bf linearly separable} if there exist a ``weight vector''
$w \in \R^n$ (viewed as a column vector) and a ``threshold'' $\th \in \R$
such that $f(x) = x w - \th$ serves as a discriminant function.
Equivalently, the data set is linearly separable if there exist a
weight vector $w \in \R^n$ and a threshold $\th \in \R$ such that
\bd
x^i w > \th \fa i \in \M_1 , x^i w < \th \fa i \in \M_2 .
\ed
To put it yet another way, given a weight $w$ and a threshold $\th$, define
$\H = \H(w,\th)$ by
\bd
\H := \{ x \in \R^n : x w - \th = 0 \} .
\H_+ := \{ x \in \R^n : x w - \th > 0 \} ,
\H_- := \{ x \in \R^n : x w - \th < 0 \} .
\ed
The data set is linearly separable if there exists a hyperplane $\H$ such that
$x^i \in \H_+ \fa i \in \M_1$, and $x^i \in \H_- \fa i \in \M_2$.

The situation can be depicted as in Figure \ref{fig:a},
where the vermilion dots denote vectors in Class $\M_1$ and the dark
green dots denote vectors in Class $\M_2$; however, it is clear that
linear separability is not affected by swapping the class labels.

\bfig[h]

\bc

\btp[line width=2pt]


\draw [dashed] (0,0) -- (6,3) ;


\filldraw [verm] (2,0) circle (2pt) ;
\filldraw [verm] (3,1) circle (2pt) ;
\filldraw [verm] (4,1) circle (2pt) ;
\filldraw [verm] (5,1) circle (2pt) ;
\filldraw [verm] (5,2) circle (2pt) ;
\filldraw [verm] (6,2) circle (2pt) ;


\filldraw [bggreen] (0,1) circle (2pt) ;
\filldraw [bggreen] (1,2) circle (2pt) ;
\filldraw [bggreen] (1,3) circle (2pt) ;
\filldraw [bggreen] (1.5,2.5) circle (2pt) ;
\filldraw [bggreen] (2,3) circle (2pt) ;
\filldraw [bggreen] (2.5,2.5) circle (2pt) ;
\filldraw [bggreen] (4,3) circle (2pt) ;
\filldraw [bggreen] (3.5,3) circle (2pt) ;

\etp

\ec

\caption{A Linearly Separable Data Set}

\label{fig:a}

\efig

It is easy to determine whether or not a given data set is linearly
separable, because that is equivalent to the feasibility of a linear
programming problem.
More general results can be obtained using Vapnik-Chervonenkis theory;
see \cite{Wenocur-Dudley81,MV03}.
Suppose that the $n$ vectors $x_1 , \ldots , x_n$ do not lie on an
$(p-1)$-dimensional hyperplane in $\R^p$.
In such a case, whenever $p \geq n-1$, the data set is linearly separable
for {\it every one\/} of the $2^n$ ways of assigning labels to the $n$ vectors.
This result suggests that, if a given data set is not linearly separable,
it can be made so by increasing the dimension of the data vectors $x^i$,
for instance, by including not just the original components but also their
higher powers.
This is the rationale behind so-called ``higher-order'' SVMs, or more
generally, kernel-based classifiers; see e.g.\
\cite{Cristianini-Shawe_Taylor00,Schollkopf-Smola02}.

It is also easy to see that, if there exists {\it one\/} hyperplane
that separates the two classes, there exist {\it infinitely many\/} such
hyperplanes.
The Support Vector Machine (SVM) introduced in \cite{Cortes-Vapnik97}
chooses the separating hyperplane
such that {\it the nearest point to the hyperplane within each class
is as far as possible from it}.
In the original SVM formulation, the distance to the hyperplane is
measured using the Euclidean or $\ell_2$-norm.
To illustrate the concept, the same data set as in Figure \ref{fig:a}
is shown again in Figure \ref{fig:b},
with the ``optimal'' separating hyperplane, and
the closest points to it within the two classes shown as hollow circles.

\bfig[h]

\bc

\btp[line width=2pt]


\draw [dashed] (0,-0.2) -- (6,3.5) ;


\filldraw [verm] (2,0) circle (2pt) ;
\draw [verm] (3,1) circle (2pt) ;
\filldraw [verm] (4,1) circle (2pt) ;
\filldraw [verm] (5,1) circle (2pt) ;
\filldraw [verm] (5,2) circle (2pt) ;
\filldraw [verm] (6,2) circle (2pt) ;


\filldraw [bggreen] (0,1) circle (2pt) ;
\filldraw [bggreen] (1,2) circle (2pt) ;
\filldraw [bggreen] (1,3) circle (2pt) ;
\filldraw [bggreen] (1.5,2.5) circle (2pt) ;
\filldraw [bggreen] (2,3) circle (2pt) ;
\filldraw [bggreen] (2.5,2.5) circle (2pt) ;
\draw [bggreen] (4,3) circle (2pt) ;
\filldraw [bggreen] (3.5,3) circle (2pt) ;

\etp

\ec

\caption{Optimal Separating Hyperplane}

\label{fig:b}

\efig

In symbols, the SVM is obtained by solving the following optimization problem:
\bd
\max_{w,\th} \min_i \inf_{v \in \H} \nm v - x^i \nm .
\ed
An equivalent formulation of the SVM is obtained by observing that
the distance of the separating hyperplane to the nearest points is
given by $c/\nm w \nm$, where
\bd
c := \min_{i \in \M_1} | y_i ( x^i w - \th) |
= \min_{i \in \M_2} | y_i ( x^i w - \th) | ,
\ed
where the equality of the two terms follows from the manner in which
the separating hyperplane is chosen.
Moreover, the optimal hyperplane is invariant under scale change, that is,
multiplying $w$ and $\th$ by a positive constant.
Therefore there is no loss of generality in taking the constant $c$
to equal one.
With this rescaling, the problem at hand becomes the following:
\be\label{eq:41}
\min_{w} \nm w \nm \st x^i w \geq 1 \fa i \in \M_1 ,
x^i w \leq -1 \fa i \in \M_2 .
\ee
This is the manner in which the SVM is implemented nowadays in most
software packages.

If the norm in (\ref{eq:41}) is the $\ell_2$-norm, then the minimization
problem (\ref{eq:41}) is a quadratic programming problem,
which can be solved
efficiently for extremely large data sets.
Moreover,
the introduction of new data points does not alter the optimal
hyperplane, unless one of the new data points is closer to the hyperplane
than the earlier closest points.
This is illustrated in Figure \ref{fig:c}, which contains exactly the
same vectors as in Figure \ref{fig:b}, plus two more shown in blue
and red respectively.
The optimal hyperplane remains the same.
For all these reasons,
the SVM offers a very attractive approach to finding a classifier
in situations where the number of features is smaller than the number of
samples.

\bfig[h]

\bc

\btp[line width=2pt]


\draw [dashed] (0,-0.2) -- (6,3.5) ;


\filldraw [verm] (2,0) circle (2pt) ;
\draw [verm] (3,1) circle (2pt) ;
\filldraw [verm] (4,1) circle (2pt) ;
\filldraw [verm] (5,1) circle (2pt) ;
\filldraw [verm] (5,2) circle (2pt) ;
\filldraw [verm] (6,2) circle (2pt) ;


\filldraw [bggreen] (0,1) circle (2pt) ;
\filldraw [bggreen] (1,2) circle (2pt) ;
\filldraw [bggreen] (1,3) circle (2pt) ;
\filldraw [bggreen] (1.5,2.5) circle (2pt) ;
\filldraw [bggreen] (2,3) circle (2pt) ;
\filldraw [bggreen] (2.5,2.5) circle (2pt) ;
\draw [bggreen] (4,3) circle (2pt) ;
\filldraw [bggreen] (3.5,3) circle (2pt) ;


\filldraw [blue] (1,2) circle (2pt) ;
\filldraw [red] (3,0.2) circle (2pt) ;

\etp

\ec

\caption{Optimal Separating Hyperplane}

\label{fig:c}

\efig

Unfortunately, in biological applications, the situation is usually the
reverse:
The number of features (the dimension of the vectors $x^i$) is a few
orders of magnitude larger than the number of samples (the number of
vectors).
In such a case, because of the results in \cite{Wenocur-Dudley81},
linear separability is not an issue.
However, in general, {\it every component\/} of the optimal weight vector $w$
is nonzero.
This means that a classifier uses every single feature in order to
discriminate between the classes.
Clearly this is undesirable.

\subsection{The $\ell_1$-Norm Support Vector Machine}\label{ssec:42}

In this subsection, we introduce a modification of the original SVM
formulation due to \cite{Bradley-Mangasarian98}.
At the same time, we also incorporate a further feature for
trading off the false positive error rate and the false negative error rate,
due originally to \cite{Veropoulos-et-al99}.

The original SVM formulation presupposes that the data set is linearly
separable.
This naturally raises the question of what is to be done in case the
data set is {\it not\/} linearly separable.
One way to approach the problem is to choose a hyperplane that
misclassifies the fewest number of points.
While appealing, this approach is impractical, because it is known
that this problem is NP-hard; see \cite{Hoffgen-et-al95,Natarajan95}.
An alternate approach is to formulate a convex relaxation of this NP-hard
problem by introducing slack variables into the
constraints in (\ref{eq:41}), and then minimizing an appropriate norm
of the vector of slack variables.
If we choose a particular norm $\nm \cdot \nm$ to measure
distances in ``feature space'',
then distances in ``weight space'' should be measured using the so-called
{\bf dual norm}, defined by
\bd
\nm w \nm_d := \sup_{\nm x \nm \leq 1} | x w | .
\ed
In particular, if we measure distances in feature space using the $\ell_1$-norm,then distances in weight space should be measured using its dual, which is
the $\ell_\infty$-norm.
With this observation, the problem can be formulated as follows:
\bd
\min_{w , \th, y, z} 
(1 - \l) \left[ \sum_{i=1}^{m_1} y_i + \sum_{i=1}^{m_2} z_i \right]
+ \l \max_{1 \leq i \leq n} | w_i | \st
\ed
\bd
x^i w - \th + y_i \geq 1 \fa i \in \M_1 ,
x^i w - \th - z_i \leq -1 \fa i \in \M_2 ,
\ed
\be\label{eq:43}
y \geq \bz_{m_1} , z \geq \bz_{m_2} .
\ee
This can be converted to
\bd
\min_{w , \th, y, z}  
(1 - \l) \left[ \sum_{i=1}^{m_1} y_i + \sum_{i=1}^{m_2} z_i \right] 
+ \l v \st
\ed 
\bd
x^i w - \th + y_i \geq 1 \fa i \in \M_1 ,
x^i w - \th - z_i \leq -1 \fa i \in \M_2 ,
\ed
\be\label{eq:44}
y \geq \bz_{m_1} , z \geq \bz_{m_2} ,
v \geq w_i \fa i  , v \geq - w_i \fa i .
\ee
This is clearly a linear programming problem.
In this formulation, $\l$ is a ``small'' constant in $(0,1)$, much closer to
$0$ than it is to $1$.
Suppose that the original data set is linearly separable, and let $w^*,\th^*$
denote a solution to the optimization problem in (\ref{eq:41}),
where $\nm w \nm_d$ replaces $\nm w \nm$.
Then the choice
\bd
w = w^*, \th= \th^* , y = \bz_{m_1} , z = \bz_{m_2} 
\ed
is certainly {\it feasible\/} for the optimization problem (\ref{eq:43}).
Moreover, if $\l$ is sufficiently small, any reduction in $\nm w \nm_d$
achieved by violating the linear separation constraints (i.e., permitting
some $y_i$ or $z_i$ to be positive rather than zero)
is offset by the increase in the term
$(1 - \l) \nm (y , z) \nm$.
It is therefore clear that, if the data set is linearly separable,
then there exists a critical value $\l_0 > 0$
such that, for all $\l < \l_0$, the optimization problem (\ref{eq:43})
has $(w^*,\th,\bz_{m_1},\bz_{m_2})$ as a solution.
On the other hand, the optimization problem (\ref{eq:43})
remains meaningful even when the data is not linearly separable.

The final aspect of the problem, as suggested in \cite{Veropoulos-et-al99},
is to introduce a trade-off between false positives and false negatives.
In this connection, it is worthwhile to recall the definitions
of the accuracy etc.\ of a classifier.
Given a discriminant function $f(\cdot)$, define
\bd
\C_1 := \{ i \in \M : f(x^i) > 0 \} ,
\C_2 := \{ i \in \M : f(x^i) < 0 \} .
\ed
Thus $\C_1$ consists of the samples that are assigned to Class 1
by the classifier, while $\C_2$ consists of the samples that are
assigned to Class 2.
Then this leads to the array shown below:
\bd
\ba{ccc}
& \C_1 & \C_2 \\ \M_1 & TP & FN \\ \M_2 & FP & TN \ea
\ed
In the above array, the entries $TP, FN, FP, TN$ stand for ``true positive'',
`` false negative'', ``false positive'' and ``true negative'' respectively.

\begin{definition}\label{def:71}
With the above definitions, we have
\be\label{eq:71}
Se = \frac{ TP }{ TP + FN } = \frac{ | \C_1 \cap \M_1 | } { | \M_1 | } ,
\ee
\be\label{eq:72}
Sp = \frac{ TN } { FP + TN } = \frac{ | \C_2 \cap \M_2 | } { | \M_2 | } ,
\ee
\be\label{eq:73}
Ac = \frac{ TP + TN }{ TP + TN + FP + FN }
= \frac{ | \C_1 \cap \M_1 | + | \C_2 \cap \M_2 | } { | \M_1 | + | \M_2 | } ,
\ee
where $Se,Sp,Ac$ stand for the {\bf sensitivity}, {\bf specificity},
and {\bf accuracy} respectively.
\end{definition}

All three quantities lie in the interval $[0,1]$.
Moreover, accuracy is a convex combination of sensitivity and specificity.
In particular,
\bd
Ac = Se \cdot \frac{ | \M_1 | } { | \M_1 | + | \M_2 | }
+ Sp \cdot \frac{ | \M_2 | } { | \M_1 | + | \M_2 | } .
\ed
Therefore
\bd
\min \{ Se, Sp \} \leq Ac \leq \max \{ Se, Sp \} .
\ed
Also, the accuracy of a classifier will be roughly equal to the sensitivity
if $\M_1$ is far larger than $\M_2$, and roughly equal to the specificity
if $\M_2$ is far larger than $\M_1$.

In many classification problems the consequences of misclassification
are not symmetric.
To capture these kinds of considerations, another parameter $\al \in (0,1)$
is introduced, and the objective function in the optimization problem
(\ref{eq:43}) is modified by making the substitution
\bd
\sum_{i=1}^{m_1} y_i + \sum_{i=1}^{m_2} z_i
\leftarrow \al \sum_{i=1}^{m_1} y_i + (1 - \al) \sum_{i=1}^{m_2} z_i ,
\ed
where we adopt the computer science notation $\leftarrow$ to mean
``replaces.''
If $\al = 0.5$, then both false positives and false negatives are weighted
equally.
If $\al > 0.5$, then there is greater emphasis on correctly classifying
the vectors in $\M_1$, and the reverse if $\al < 0.5$.
%
%
With this final problem formulation, the following desirable properties result:
\bit
\item The problem is a linear programming problem and is therefore tractable
for even for extremely large values of $n$, the number of features.
\item The formulation can be applied without knowing beforehand whether or
not the data set is linearly separable.
\item The formulation provides for a trade-off between false positives
and false negatives.
\item Most important, the optimal weight vector $w$ has at most $m$
nonzero entries, where $m$ is the number of samples.
Hence the classifier uses at most $m$ out of the $n$ features.
\eit
For these reasons, the $\ell_1$-norm SVM forms the starting point for
our further research into classification.

\subsection{The Lone Star Algorithm}\label{ssec:lone-star}

As pointed out in Section \ref{sec:4}, both the traditional $\ell_2$-norm
support vector machine (SVM) as well as the $\ell_1$-norm SVM can
be used for two-class classification problems.
When the number of samples $m$ is far larger than the number of features $n$,
the traditional SVM performs very satisfactorily, whereas the $\ell_1$-norm
SVM of \cite{Bradley-Mangasarian98} is to be preferred when $m < n$.
Moreover, the $\ell_1$-norm SVM is guaranteed to use no more than $m$
features.
However, in many biological applications, even $m$ features are too many.
Biological measurements suffer from poor repeatability.
Therefore a classifier that uses fewer features would be far preferable
to one that uses more features.
In this section we present a new algorithm for two-class classification
that often uses far fewer than $m$ features, thus making it very
suitable for biological applications.
The algorithm combines the $\ell$-norm SVM of \cite{Bradley-Mangasarian98},
recursive feature elimination (RFE) of \cite{Guyon-et-al02},
and stability selection of \cite{Mein-Buhl10}.
A preliminary version of this algorithm was reported in \cite{MV-CDC12a}.

The algorithm is as follows:
\ben
\item
Choose at random a ``training set'' of samples of size $k_1$ from $\M_1$
and size $k_2$ from $\M_2$, such that $k_l \leq m_l/2$, and $k_1,k_2$
are roughly equal.
Repeat this choice $s$ times, where $s$ is a ``large'' number.
This generates $s$ different ``training sets'', each of which consists of
$k_l$ samples from $\M_l$, $l = 1, 2$.
\item
For each randomly chosen training set, compute a corresponding
 optimal $\ell_1$-norm SVM using the formulation (\ref{eq:43}).
This results in $s$ different optimal weight vectors and thresholds.
\item Let $k$ denote the average number of nonzero entries in the
optimal weight vector across all randomized runs.
Average all $s$ optimal weight vectors and thresholds,
retain the largest $k$ components
of the averaged weight vector and corresponding feature set,
and set the remaining components to zero.
This results in reducing the number of features from the original $n$ to $k$.
\item Repeat the process with the reduced feature set, 
but the originally chosen randomly selected training samples,
until no further reduction is possible in the number of features.
This determines the final set of features to be used.
\item
Once the final feature set is determined,
carry out two-fold cross validation by dividing the data $s$ times into
a training set of $k_1 , k_2$ randomly selected samples and assessing
the performance of the resulting $\ell_1$-norm classifier on the
testing data set, which is the remainder of the samples.
Average the weights generated by the $t \leq s$ best-performing classifiers
where $t$ is chosen by the user, and call that the final classifier.
\een

When the number of features $n$ is extremely large, an optional
pre-processing step is to compute the mean value of each of the $n$
features for each class, and retain only those features wherein
the difference between means is statistically significant using
the ``Student'' $t$-test.
Our experience is that using this optional pre-processing step does
not change the final answer very much, but does decrease the CPU time
substantially.

Now some comments are in order regarding the above algorithm.
\bit
\item
In some applications,
$\M_1$ and $\M_2$ are of comparable size, so that the size of the training
set can be chosen to equal roughly half of the total samples within each class.
However, in other applications, the sizes of the two sets are dissimilar,
in which case the larger set has far fewer of its samples used in training.
\item
Step 1 of randomly choosing $s$ different training sets differs from
\cite{Guyon-et-al02}, where there is only one randomized division
of the data into training and testing sets.
\item
For each random choice of the training set,
the {\it number\/} of nonzero entries in the optimal weight vector
is more or less the same; however, the {\it locations\/} of nonzero entries
in the optimal weight vector vary from one run to another.
\item
In Step 3 above, instead of averaging the optimal weights over all $s$
runs and then retaining the $k$ largest components, it is possible
to adopt another strategy.
Rank all $n$ indices in order of the number of times that index has a nonzero
weight in the $s$ randomized runs, and retain the top $k$ indices.
In our experience, both approaches lead to virtually the same choice
of the indices to be retained for the next iteration.
\item Instead of choosing $s$ randomized training sets right at the outset,
it is possible to choose $s$ randomized training sets each time the
number of features is reduced.
\item
In the final step,
there is no distinction between the training and testing
data sets, so the final classifier is run on the entire data set to
arrive at the final accuracy, sensitivity and specificity figures.
\eit

The advantage of the above approach vis-a-vis the $\ell_2$-norm SVM-RFE
of \cite{Guyon-et-al02}
is that the number of features reduces significantly at each step,
and the algorithm converges in just a few steps.
This is because, in the $\ell_1$-norm SVM, many components of the weight
vector are ``naturally'' zero, and need not be truncated.
In contrast, in general all the components of the weight vector
resulting from the $\ell_2$-norm SVM will be nonzero; as a result
the features can only be eliminated one at a time, and in general
the number of iterations is equal to (or comparable to) $n$, the
initial number of features.

The new algorithm can be appropriately referred to as the 
``$\ell_1$-SVM $t$-test and RFE'' algorithm, where SVM and RFE
are themselves acronyms as defined above.
Once again taking the first letters, we are led to the ``second-level''
acronym ``$\ell_1$-StaR'', which can be pronounced as ``ell-one star''.
Out of deference to our domicile, we have decided to call it
the ``lone star'' algorithm.

The lone star algorithm was applied to the problem of predicting which
patients of endometrial cancer are at risk of lymph node metastasis.
These results are reported elsewhere.
But in brief the situation is the following:
The endometrium is the lining of the uterus.
When a patient contracts endometrial cancer, her
uterus, ovaries, and fallopian tubes are surgically removed.
One of the major risks run by endometrial cancer patients is
that the cancer will metastasize and spread through the body via
pelvic and/or para-aortic lymph nodes.
The Gynecological Oncology Group (GOG) recommends that the patient's
pelvic and para-aortic lymph nodes should also be surgically removed
when the size of the tumor exceeds 2cm in diameter.
However, post-surgery analysis reveals that even in this case,
lymphatic metastasis is present in only 22\% of the cases
\cite{Mariani-et-al08}.

To predict the possibility of lymphatic metastasis, 1,428 micro-RNAs
were extracted from 94 tumors, half with and half without metastasis.
Using the lone star algorithm, 13 micro-RNAs were identified as
being highly predictive.
When tested on the entire training sample of 94 tumors, the lone star
classifier correctly classified 41 out of 43 lymph-positive samples,
and 40 out of 43 lymph-negative samples.
In on-going work, these micro-RNAs were measured on an independent
cohort of 19 lymph-negative and 9 lymph-positive tumors.
The classifier classified 8 out of 9 lymph-positive tumors correctly,
and 11 out of 19 lymph-negative tumors correctly.
Thus, while the specificity is not very impressive, the sensitivity
is extremely good, which is precisely what one wants in such a situation.
Moreover, using a two-table contingency analysis and the Barnard exact
test, the likelihood of arriving at this assignment by pure chance
(the so-called $P$-value) is bounded by $0.011574$.
In biology any $P$-value less than $0.05$ is generally considered 
to be significant.

\section{Some Topics for Further Research}\label{sec:9}

Machine learning and computational biology
are both vast subjects, and their intersection contains many more topics
than are touched upon in this brief article.
Besides, there are other topics in computational cancer biology
that do not naturally belong to machine learning, for example,
modelling tumor growth using branching processes.
Therefore the emphasis in this article has been on topics that are
well-established in the machine learning community, and are also
relevant to problems in computational cancer biology.

Until now several ``penalty'' norms have been proposed for
inducing an optimization algorithm to select
structured sparse feature sets, such as group lasso (GL) and
sparse group lasso (SGL).
As pointed out in Section \ref{sec:5}, available extensions of these
penalty norms to overlapping sets do not address biological
networks where there are multiple paths from a master regulator
to a final node.
Any advance in this direction would have an immediate application
to computational biology.

Compressed sensing theory as discussed in Section \ref{sec:6} 
is based on the premise that {\it is possible to choose\/}
the measurement matrix $A$.
The available theorems in this theory are based on assumptions on
the measurement matrix, such as the restricted isometry property,
or the null space property,
and perhaps something even more general in future.
In order to apply techniques from compressed sensing theory to
cancer biology, it would be necessary to modify the theory to the case
where the measurement matrix is given, and not chosen by the user.
The RIP corresponds to the assumption that in an $m \times n$ matrix $A$,
every choice of $k$ columns results in a nearly orthogonal set.
In actual biological data, such an assumption has no hope of being true,
because the expression levels of some genes would be highly correlated
with those of other genes.
In \cite{Candes-Plan09}, the authors suggest that it is possible to handle
this situation by first clustering the column vectors and then choosing
just one exemplar from each cluster before applying the theory.
Our preliminary attempts to apply such an approach to ovarian cancer data
(\cite{TCGA-Ovarian}) are not very promising, leading to RIP orders of
5 or 10 -- far too small to be of practical use.
Thus there is a need for the development of other heuristics besides
clustering to extract nearly orthogonal sets of columns for actual
measurement matrices.
In this connection it is worth pointing out \cite{Huang-Zhang10}
that group RIP is easier to achieve using random projections, as
compared to RIP.
However, it is not clear whether a ``given'' $A$ matrix is likely
to satisfy a group RIP with a sufficiently large order.

In general it would appear that sparse regression is more advanced
than sparse classification, with both well-established theoretical foundations
as well as widely used algorithms in the former.
In contrast, sparse classification does not have such a wealth of results.
The lone star algorithm introduced here has performed well in several
applications involving cancer data, and at least for the moment,
it appears to be the only available method to select far fewer features
than the size of the training set of samples.
As of now there is no theoretical justification for this observed
behavior.
Recall that the $\ell_1$-norm SVM is guaranteed only to choose no more
features than the size of the training set; but there is no reason
to assume that it will use fewer.
Therefore it is certainly worthwhile to study when and why lone star
and other such algorithms will prove to be effective.

\section*{Acknowledgement}

The author would like to assert his deep gratitude to Professor
Michael A.\ White of the UT Southwestern Medical Center, Dallas, TX
for introducing him to the fascinating world of cancer biology and
for turning him into a passable imitation of a computational cancer
biologist.
He would also like to thank his students Eren Ahsen, Burook Misganaw
and Nitin Singh for carrying out much of the work that supports the theory
reported here.

\bibliographystyle{natbib}

\bibliography{Cancer-Refs,Comp-Sens}

\end{document}